\documentclass[a4paper,12pt,times new roman]{article}

\usepackage[top=1in, bottom=1in, left=24mm, right=24mm]{geometry}
\usepackage{graphicx}
\usepackage{epsfig}

\begin{document}
\vspace{10pt}
\title{\textbf{Dynamics of ion cloud in a linear Paul trap}} \vspace{5pt}
\author{\bf{Pintu Mandal$^{\ast}$, Manas Mukherjee$^{\dag}$} \\
\\
Raman Center for Atomic, Molecular and Optical Sciences \\
Indian Association for the Cultivation of Science \\
2A \& 2B Raja S. C. Mullick Road, Kolkata 700 032 \\
\\
$^{\dag}$ \emph{Present address:} Centre for Quantum Technologies \\
National University of Singapore, Singapore - 117543 \\
\\
$^{\ast}$Email: \emph{pintuphys@gmail.com}} \maketitle \vspace{5pt}
\begin{center}
\textbf{Abstract} \\
\end{center}
A linear ion trap setup has been developed for studying the dynamics
of trapped ion cloud and thereby realizing possible systematics of a
high precision measurement on a single ion within it. The dynamics
of molecular nitrogen ion cloud has been investigated to extract the
characteristics of the trap setup. The stability of trap operation
has been studied with observation of narrow nonlinear resonances
pointing out the region of instabilities within the broad stability
region. The secular frequency has been measured and the motional
spectra of trapped ion oscillation have been obtained by using
electric dipole excitation. It is applied to study the space charge
effect and the axial coupling in the radial plane.

\newpage

\section{Introduction}\label{section1}

``A single atomic particle forever floating at rest in free
space"~\cite{Dehmelt88} is an ideal system for precision measurement
and a single trapped ion provides the closest realization to this
ideal. A single or few ions can be trapped within a small region of
space in an ion trap and they are free from external perturbations.
Such a system has been used for the precision measurement of
electron's g - factor~\cite{Dyck87}, various atomic properties like
the lifetime of atomic states~\cite{Yu97}, the quadrupole
moment~\cite{Barwood04,Oskay05,Roos06} \emph{etc}. Precision
table-top experiments of fundamental physics in the low energy
sector like the atomic parity violation measurement, nuclear anapole
moment measurement, electron's electric dipole moment measurement
are either in progress in different laboratories worldwide or
proposed~\cite{Fortson93,Mandal10,Sahoo11,Versolato10}. Any
high-precision experiment appears with systematics which are
required to be tracked or removed and hence a systematic
investigation on the system itself is essential at the initial
stage. In order to prepare for measuring atomic parity violation
with trapped ions, a series of experiments have been performed in a
linear ion trap to fully understand its behaviour and associated
systematics. In this colloquium, the results of some experiments
will be presented that are of preeminent interest to an audience
coming from a variety of physics disciplines. It is organised with a
brief overview on the physics of ion trapping in a linear Paul trap,
description of the experimental setup and followed by results.

\section{Physics of ion trapping}\label{section2}

\begin{figure}[hh]
\vspace{-1cm}\hspace{-1.5cm}
  \includegraphics[width=0.9\textwidth, angle=270]{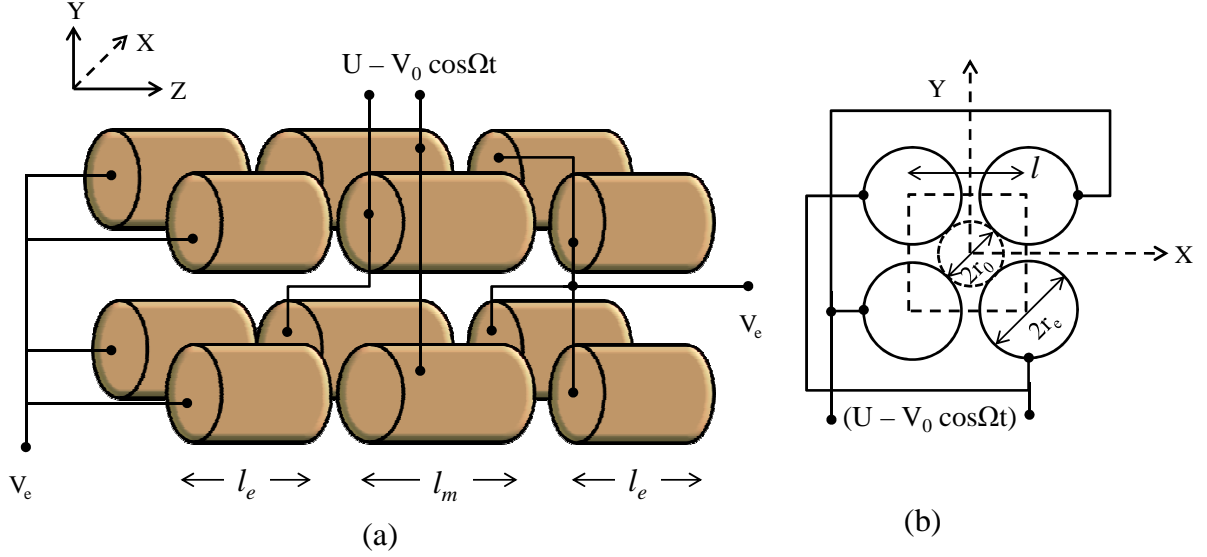}\\
 \vspace{-6cm}
  \caption{(a) Schematic of the linear ion trap used in the experiment. (b) End view of the
  four middle electrodes with relevant electrical connections. Various dimensions as
  marked by $l_{e}$, $l_{m}$, $l$, $r_{e}$ and $r_{0}$ are described in
  section~\ref{section3}. }
  \label{lintrap}
\end{figure}

An electrostatic field can not produce a potential minimum in three
dimensional space as is required for trapping the charged particles.
It is therefore, either a combination of static magnetic field and
an electric field is used (Penning trap) or a combination of a
time-varying and an electrostatic field is used (Paul trap). In Paul
trap a radio-frequency (rf) potential superposed with a dc potential
is applied on electrodes of hyperbolic geometry to develop
quadrupolar potential in space. The geometry of the electrodes
evolved over the decades for ease in machining, smooth optical
access to the trapped ions \emph{etc}. Figure~\ref{lintrap} shows
one of the most frequently used trap geometries with four
three-segmented rods placed symmetrically at four corners of a
square and is commonly called a linear Paul trap. The four rods at
each end are connected together and a common dc potential ($V_{e}$)
is applied so as to produce an axial trapping potential. The
diagonally opposite rods at the middle are connected and a rf
($V_{0}\cos\Omega t$) in addition to a dc potential ($U$) is applied
on one pair with respect to the other pair for providing a dynamic
radial confinement. The radial potential inside the trap is
\begin{equation}\label{eqn1}
\Phi(x,y,t) = (U-V_{0}\cos\Omega
t)\left(\frac{x^{2}-y^{2}}{2r_{0}^{2}}\right),
\end{equation}
where $2r_{0}$ is the separation between the surfaces of the
diagonal electrodes as depicted in figure~\ref{lintrap}(b). The
equipotential lines are rectangular hyperbolae in the $xy$ plane
having four-fold symmetry about the $z$ axis. The equation of motion
of an ion of charge $e$ and mass $m$ under the potential
$\Phi(x,y,t)$ (eqn.~\ref{eqn1}) can be represented as
\begin{eqnarray}\label{eqn2}
\frac{d^{2}x}{dt^{2}} &=& -\frac{e}{mr_{0}^{2}}(U-V_{0}\cos\Omega
t)x \\ \nonumber \frac{d^{2}y}{dt^{2}} &=&
\frac{e}{mr_{0}^{2}}(U-V_{0}\cos\Omega t)y.
\end{eqnarray}

These equations (eqn.~\ref{eqn2}) can be rewritten as
\begin{equation}\label{eqn3}
\frac{d^{2}u}{d\zeta^{2}} + (a_{u}-2q_{u}\cos2\zeta)u = 0,
\end{equation}
with $u=x,y$, where
\begin{eqnarray}\label{eqn4}
a_{x}&=&-a_{y}=\frac{4eU}{mr_{0}^{2}\Omega^{2}}, \nonumber \\
q_{x}&=&-q_{y}=\frac{2eV_{0}}{mr_{0}^{2}\Omega^{2}},
\end{eqnarray}
and $\zeta = \Omega t/2$. Eqn.~\ref{eqn3} is standard Matheiu
differential equation and its solution provides stability or
instability of the ion motion~\cite{Dawson76} depending on the
values of the parameters $a$ and $q$ as defined in eqn.~\ref{eqn4}.
There exists a region in $a$ vs. $q$ diagram for which the
ion-motion is stable along a particular direction, for example along
$x$. A similar stability region exists for the motion along $y$
direction. An intersection between these two stability regions thus
signifies a stable motion in $xy$ plane. For stable ion motion the
trap should be operated at $q<0.908$.

The stable solutions of Mathieu differential equation show that the
trapped ion oscillates with different frequencies given
by~\cite{Ghosh95}
\begin{equation}\label{eqn5}
\omega_{n}=\frac{(2n\pm\beta)\Omega}{2}, n = 0, 1, 2, 3...
\end{equation}
Here $\beta$ is a function of the trap operating parameters $a$, $q$
and for their small values, $\beta=\sqrt{a+q^{2}/2}$. The
fundamental frequency $\omega_{0}$ (that corresponds to $n=0$) of
secular motion and other micromotion frequencies are given by
\begin{eqnarray}\label{eqn6}
\omega_{0}&=&\frac{\beta\Omega}{2}, \\ \nonumber
\omega_{1\pm}&=&\Omega\pm\omega_{0}, \\
\omega_{2\pm}&=&\Omega\pm2\omega_{0} \nonumber
\end{eqnarray}
and so on. A large spectra of the motional frequency have been
obtained in our experiment by using electric dipole excitation
technique.

Though in ideal case the trap potential is quadrupolar, real traps
appear with misalignment, defect in machining, truncation and holes
in the electrodes to have optical access. In addition, there are
space charge developed by the trapped ions themselves. All these
result in deviation from pure quadrupole trap potential contributing
to other higher order terms and make the ion motion unstable for
certain values of the trapping parameters, for which the stability
exists in ideal case. The ions gain energy from the rf trapping
field and their motional amplitudes get enhanced resulting loss from
the trap. The condition of such nonlinear resonances is given
by~\cite{Dawson69}
\begin{equation}\label{eqn7}
n_{x}\omega_{0x}+n_{y}\omega_{0y}=\Omega,  n_{x}, n_{y}=0, 1, 2,
3...
\end{equation}
where $\omega_{0x}$ and $\omega_{0y}$ are the secular frequencies
for the motion along $x$ and $y$ respectively. Here $n_{x}+n_{y}=k$
is the order of the multipole. If one of the trap parameters is
varied, a parametric resonance appears at a definite value subjected
to the condition defined by eqn.~\ref{eqn7} and it gives rise to
instabilities called ``black canyons''~\cite{March05} within the
stability diagram.

\begin{figure}[hh]
\hspace{-1.5cm}
  \includegraphics[width=0.95\textwidth, angle=270]{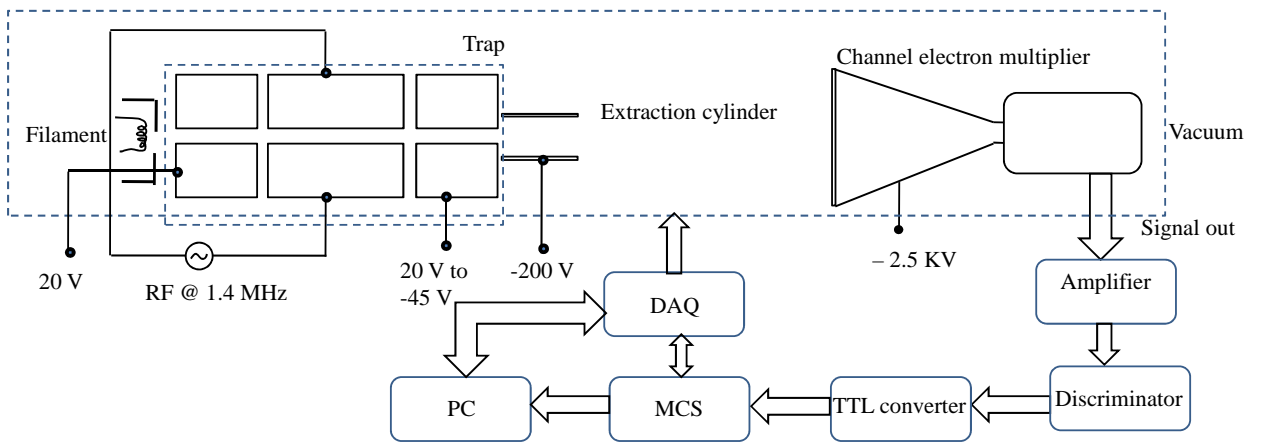}\\
  \vspace{-7.5cm}
  \caption{Schematic of the experimental setup. The trap, filament and the
  CEM with other ion optics (extraction cylinder) are housed in a vacuum chamber. The functioning
  and control of the signal processing devices are explained in the text.}
  \label{setup_schematic}
\end{figure}

\section{Experimental setup}\label{section3}

The schematic of the whole experimental setup is presented in
figure~\ref{setup_schematic}. It consists of a linear Paul trap as
shown in figure~\ref{lintrap}, an ionization setup, extraction and
detection setup. The linear trap is assembled from four
three-segmented electrodes each placed at four corners of a square
of side ($l$) $12.73$~mm~[figure~\ref{lintrap}](b). Each of twelve
rods are of diameter ($2r_{e}$) $10$~mm. The four middle rods are of
length ($l_{m}$) $25$~mm while all others are $15$~mm long
($l_{e}$)~[figure~\ref{lintrap}(a)]. The separation between the
surfaces of the diagonally opposite rods ($2r_{0}$) is $8$~mm. The
middle electrode is separated from the end electrodes by a gap of
$2$~mm. The molecular nitrogen ions (N$_{2}^{+}$) are created by
electron impact ionization. The ions are dynamically trapped for few
hundreds ms before they are extracted by lowering the axial
potential in one direction. The extracted ions are detected by a
channel electron multiplier (CEM). The CEM produces one pulse
corresponding to each ion and the pulse is successively processed
through an amplifier, a discriminator, a TTL converter before it is
fed into a multichannel scalar (MCS) card which ultimately counts
the number of ions reaching the CEM. This time-of-flight (TOF)
technique provides a detection efficiency around $10\%$. The time
sequences are generated by National Instruments' Data Acquisition
(DAQ) hardware which is controlled by Labview and monitored by a
personal computer (PC).

The trap is operated at a rf frequency of $1.415$~MHz and no dc
potential is applied to the middle electrodes ($U=0$, $a_{u}=0$).
The end electrodes are kept at $+20$~V while trapping. At the time
of extraction, the end electrodes at the ion-exit-side are switched
fast (within $75~ns$) from $+20$~V to $-45$~V.

\section{Experimental results}\label{section4}

\begin{figure}[hh]
\hspace{1.2cm}
  \includegraphics[width=0.6\textwidth, angle=270]{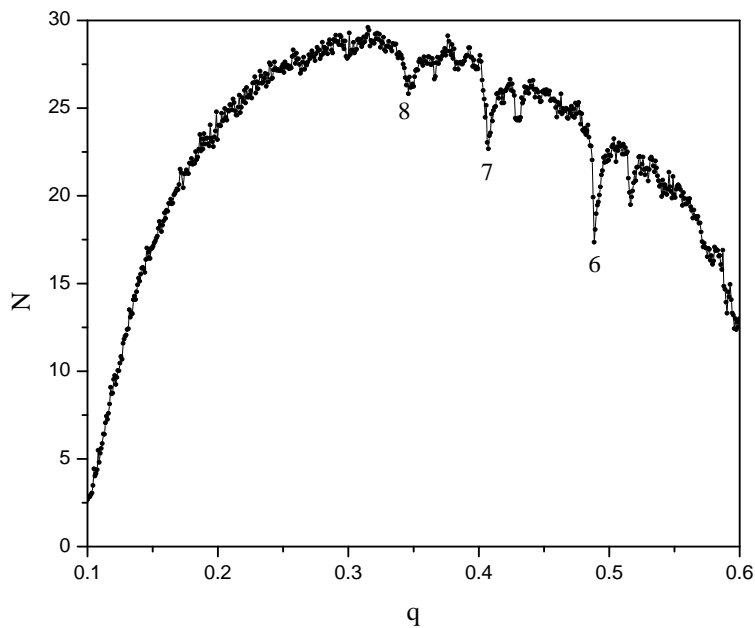}\\
  \caption{Number of trapped ions ($N$) as a
  function of $q$ ($a=0$). Sudden fall of $N$
  about some specific values of $q$ corresponds to nonlinear resonances as explained in the text.
  The numbers $6$, $7$, $8$ describe the order of the multipoles to which the resonances are assigned.}
  \label{q_scan}
\end{figure}

\subsection{Stability characteristics}

The stability behaviour of the trap is studied by varying the
trap-operating-parameter $q$ while keeping the other parameter $a$
at zero. The $q$ is varied in steps of $0.0008$ by changing the rf
amplitude at small intervals of $0.35$~V while the number of trapped
ions ($N$) is plotted in figure~\ref{q_scan} as a function of $q$.
It shows that $N$ grows with $q$ initially but decreases above
$q\approx0.5$. It remains almost constant and shows a plateau for
$0.3<q<0.5$. The q scanning is restricted to $0.6$ due to the
presence of heavier masses which can not be resolved in the TOF
spectra.

One of the significant observations within this stability diagram is
the appearance of narrow nonlinear resonances for specific values of
$q$. These are due to the existence of higher order multipoles
within the trap potential as explained in section~\ref{section2}.
The resonances appear at $q=0.3461, 0.4073$ and $0.4885$ are
assigned to the $8$th,$7$th and $6$th order multipoles respectively.
The $7$th order multipole is unlikely as the symmetry of trap setup
forbids non-zero perturbations due to odd order multipole. However,
such a nonlinear observation has been observed
previously~\cite{Drakoudis06}. It could result from any misalignment
of the setup that partially breaks the radial symmetry or due to
some electrical connection wires near the trap center. The nonlinear
resonance at $q=0.5163$ in our experiment could not be assigned. It
may result from other atmospheric species, or some molecules
produced by charge-transfer-reactions inside the trap. As can be
seen from figure~\ref{q_scan}, the depth of the resonance appearing
at $q=0.4885$ is maximum and hence it can be concluded that the
$6$th order multipole is the strongest one in our trap setup.

\begin{figure}[hh]
  \includegraphics[width=0.75\textwidth, angle=270]{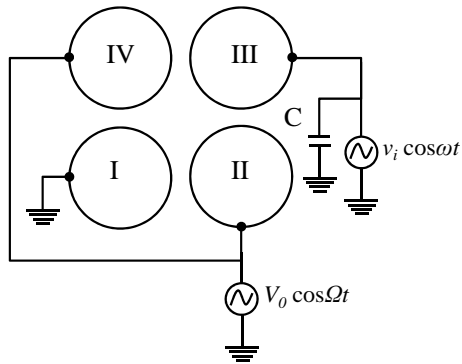}\\
  \vspace{-6cm}
  \caption{Schematic of the circuit used for dipole excitation of trapped ions. The dipole excitation signal
  $v_{i}\cos\omega t$ is
  applied between the electrodes marked as I and III.}
  \label{circuit_dipole}
\end{figure}

While operating the trap for a single ion, the region of
instabilities should be avoided as the ion gains energy from the
time varying trapping field corresponding to these operating regions
and its motional amplitude increases. It can add to systematics in
precision measurement on the ion.

\subsection{Dipole excitation of trapped ions}

Electric dipole excitation of the trapped ions has been employed to
measure their secular frequency and to obtain motional spectra. An
electric dipole field has been applied on one of the middle
electrodes as shown schematically in figure~\ref{circuit_dipole}.
The amplitude of the excitation potential ($v_{i}$) is kept fixed
while its frequency is tuned so as to match with the secular
frequency of the trapped ions. The trap operating parameters are
kept fixed during the experiment. After the ions are loaded into the
trap, the dipole excitation field is applied for few hundreds of ms.
After a short waiting time, the ions are released and detected. The
frequency of the excitation signal ($\omega$) is varied and the
total number of ions is detected in each step.

\begin{figure}[hh]
\hspace{0.5cm}
  \includegraphics[width=0.6\textwidth, angle=270]{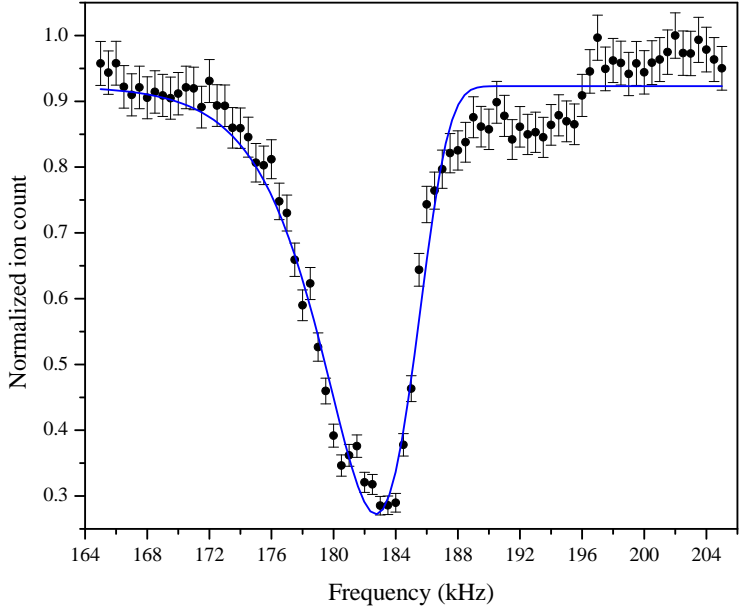}\\
  \caption{Dipole excitation resonance of trapped ions. Solid line shows a fit to the
  data with model function described in eqn.~\ref{eqn8}.}
  \label{dipole_scan}
\end{figure}

\subsubsection{Measurement of secular frequency}

The experimentally obtained ion counts ($N$) have been normalised
after dividing by the maximum ion count ($N_{max}$) during a
particular experiment. The normalised ion count ($N_{n}=N/N_{max}$)
with associated uncertainty, has been plotted as a function of the
frequency ($\omega/2\pi$) of the dipole excitation signal.
Figure~\ref{dipole_scan} shows such a dipole excitation resonance
plot obtained with an excitation amplitude $v_{i}=50$~mV. The
frequency is scanned from $165$~kHz to $205$~kHz in steps of
$500$~Hz. The excitation signal is applied during $150$~ms in each
step. The experimental data points have been fitted with the
following function,
\begin{equation}\label{eqn8}
N_{n}=N_{0}+A\exp\left[-\exp(-\omega')-\omega'+1\right],
\end{equation}
with $\omega'=(\omega-\omega_{0})/\sigma$. Here $\omega_{0}$ is the
resonant frequency and is equal to the secular frequency of the
trapped ions. $N_{0}$ is an offset, $A$ is a scaling factor and
$\sigma$ is the full-width at half-maxima (FWHM) of the resonance.
The secular frequency of the trapped ions obtained from the fit is
$182.730$($76$)~kHz and it is in good agreement with theoretically
calculated value.

\subsubsection{Motional spectra}

\begin{figure}
\vspace{-1cm} \hspace{-1.7cm}
  \includegraphics[width=0.65\textwidth, height=0.8\textheight, angle=270]{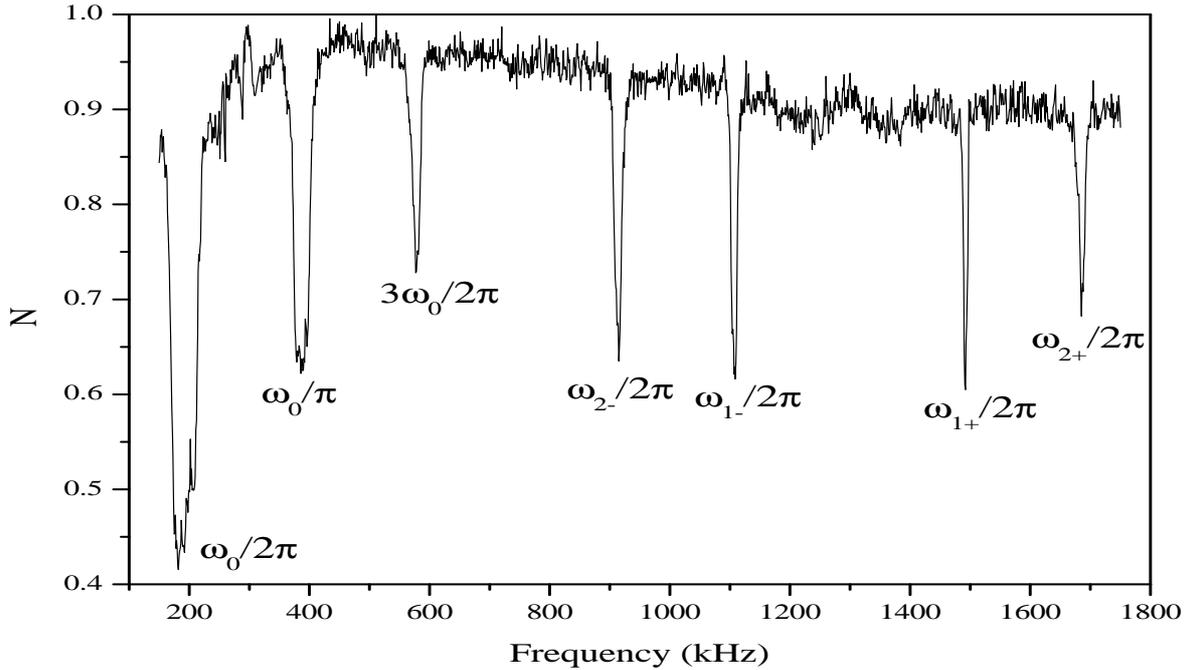}\\
  \caption{The normalized ion count $N$
  plotted as a function of the dipole excitation frequency (in kHz)
  presenting the motional spectra of the trapped ion cloud. The amplitude of the excitation voltage is
  $v_{i}=100$~mV and the trap operating parameters are set at $a=0$,
  $q=0.39$ for N$_{2}^{+}$. The frequency of the trap supply voltage is $\Omega=2\pi\times1.3$~MHz.}
  \label{motional_spectra}
\end{figure}

The motional spectra of the trapped ions as described in section~2
have been measured by varying the dipole excitation signal frequency
over long range. Figure~\ref{motional_spectra} shows the motional
spectra in the radial plane. The fundamental or the first harmonic
frequency of oscillation is observed at
$\omega_{0}=2\pi\times184$~kHz that corresponds to the trap
operating parameter $a=0$, $q=0.39$ and it is the strongest one. The
second and third harmonics are observed at $386$~kHz and $577$~kHz
respectively. The other motional spectra as described in
eqn.~\ref{eqn6} are observed at $\omega_{2-}=2\pi\times915$~kHz,
$\omega_{1-}=2\pi\times1.109$~MHz, $\omega_{1+}=2\pi\times1.492$~kHz
and $\omega_{2+}=2\pi\times1.685$~MHz.

\subsubsection{Application}

The accurate measurement of the motional frequency of the trapped
ions is essential for different studies on them~\cite{Mandal13th}.
In a real linear Paul trap the radial motion is coupled with the
axial motion and hence a variation in the axial potential affects
the secular frequency of the ions~\cite{Drakoudis06}. The motional
frequency of the trapped ions for different axial potentials has
been measured with the technique described in section~4.2.1 and from
this measurement the geometrical radial-axial coupling constant has
been determined. This is important for any precision spectroscopic
study on a single ion confined in this setup. The dipole excitation
technique is also applied to study the shift in the motional
frequency due to space charge created by the trapped ions. It is
observed that the frequency decreases while they oscillate
collectively with increasing space charge~\cite{Mandal13}. Detailed
discussion on these topics can be found
elsewhere~\cite{Mandal13th,Mandal13}.

\section{Conclusions}\label{section5}

This colloquium paper describes the development of an ion trap
facility at IACS and the results of some experiments fundamentally
based on the dynamics of a trapped ion cloud. It presents a
demonstration of some first principles of ion-trap-physics that are
of common interest to an audience coming from wide variety of
physics and participating in this colloquium. The results are also
some significant feeds to the precision measurement based on a
single ion in a linear Paul trap.

\section{Acknowledgement}\label{section6}

The authors thank S. Das, D. De Munshi and T. Dutta, presently at
the Centre for Quantum Technologies, National University of
Singapore, for their support in developing the experimental setup at
IACS, and beyond it. The machining support from Max-Planck
Institute, Germany is gratefully acknowledged.

\end{document}